\documentclass[10pt, conference, compsocconf]{IEEEtran}

\usepackage{graphicx}
\usepackage{amsmath}
\usepackage{url}

\begin{document}

\title{Building An Information System for a Distributed Testbed}

\author{
\IEEEauthorblockN{Warren Smith}
\IEEEauthorblockA{Texas Advanced Computing Center\\
University of Texas at Austin\\
10100 Burnet Road (R8700)\\
Austin, Texas 78758-4497\\
Email: wsmith@tacc.utexas.edu}
\and
\IEEEauthorblockN{Shava Smallen}
\IEEEauthorblockA{San Diego Supercomputing Center\\
9500 Gilman Drive \#505\\
La Jolla, CA 92093\\
Email: ssmallen@sdsc.edu}
}

\maketitle

\begin{abstract}

This paper describes an information system designed to support the large volume of monitoring information
generated by a distributed testbed. This monitoring information is produced by several subsystems and consists
of status and performance data that needs to be federated, distributed, and stored in a timely and easy to use
manner. Our approach differs from existing approaches because it federates and distributes information at a
low architectural level via messaging; a natural match to many of the producers and consumers of
information. In addition, a database is easily layered atop the messaging layer for consumers that want to
query and search the information. Finally, a common language to represent information in all layers of the
information system makes it significantly easier for users to consume information. Performance data shows that
this approach meets the significant needs of FutureGrid and would meet the needs of an experimental
infrastructure twice the size of FutureGrid. In addition, this design also meets the needs of existing
distributed scientific infrastructures.


\end{abstract}


\section{Introduction}

FutureGrid is a distributed testbed where users can perform experiments with cloud, grid, and high-performance
computing technologies. Distributed infrastructures like FutureGrid are complex systems that must provide
information to users and managers.  Infrastructure managers use information about the infrastructure to
determine whether it is operating correctly, to monitor usage, and to identify short- and long-term
improvements. Users need information to understand the infrastructure, to select resources and
services to use in the short- and long-term, and to monitor their usage of the infrastructure. In addition, an
experimental infrastructure like FutureGrid gathers a large amount of performance information that must be
made available to users so that they can determine how their experiments impact the infrastructure.

Heterogeneous and real-time performance information is gathered by a variety of monitoring tools and needs to
be federated and provided in an efficient and easy to use manner.  Our approach is to deploy an information
system that federates information at a low architectural level via publish/subscribe messaging. In addition,
since messaging systems typically place very few restrictions on the content of messages, our approach
specifies that a single representation language is used so that it is much easier to consume
information. Finally, while publish/subscribe messaging supports many of the use cases described in
Section~\ref{sec-usecases}, an information storage system is layered atop the messaging layer to provide a
searchable cache of recent information.

The monitoring tools available on FutureGrid are described in Section~\ref{sec-monitoring} and
Section~\ref{sec-design} describes how we use the RabbitMQ messaging service to distribute information, use
the JavaScript Object Notation (JSON) as our representation language, and use PostgreSQL to store JSON
documents in a searchable manner.

FutureGrid generates a significant volume of performance data and the information system must be able to
process it. We therefore performed experiments to ensure that our design is sufficient for the current and
possible future needs of FutureGrid. The results of our performance experiments are presented in
Section~\ref{sec-performance}. We find that FutureGrid publishes approximately 41 messages a second that this
is a rate that RabbitMQ can easily accommodate as can PostgreSQL. In addition, we find that our design could
support a large volume of custom performance monitoring information in FutureGrid and could also support
significantly larger distributed scientific computing infrastructures.

Section~\ref{sec-relatedwork} describes other information system designs and compares our approach to them. We
present conclusions and future work in Section~\ref{sec-conclusions}.

\section{FutureGrid}
\label{sec-futuregrid}

FutureGrid~\cite{futuregrid-hpcbook,futuregrid-web} is a distributed testbed funded by the National Science
Foundation that supports experiments in cloud computing, grid computing, and high performance computing
(HPC). The goals of FutureGrid are to provide a heterogeneous hardware environment, to deploy a heterogeneous
and configurable software environment, and to provide the tools so that users can perform rigorous experiments
on this infrastructure.

FutureGrid includes heterogeneous clusters at five locations in the United States. These clusters are
connected by a high-performance network and these connections pass through a network impairment device that
can introduce specified network degradations needed by experiments. Most of these clusters are
partitioned so that they simultaneously provide HPC environments and Infrastructure as a Service (IaaS)
clouds. The HPC environments consist of batch-scheduled access to nodes using Torque and Moab and the
compilers and libraries so that users can run experiments consisting of traditional serial, high-throughput,
and parallel computations. The IaaS partitions are managed by three different software systems:
Nimbus~\cite{NimbusWeb}, OpenStack~\cite{OpenStackWeb}, and Eucalyptus~\cite{EucalyptusWeb}. Multiple IaaS
infrastructures are provided so that users can evaluate and experiment with different implementations.

On top of this basic infrastructure, FutureGrid provides pre-configured virtual environments, tools for
managing distributed experiments, and a user web portal. In addition, FutureGrid provides an infrastructure
for monitoring the status and performance of FutureGrid resources and services. This monitoring infrastructure
is an important and unique feature of FutureGrid because in addition to monitoring the status of resources and
services (the type of monitoring commonly performed in distributed infrastructures), it also gathers detailed
performance information and federates all of this information into a unified system.

\section{Use Cases}
\label{sec-usecases}

Th FutureGrid monitoring information system was created to support the use cases described in this
section. Like many infrastructure projects, FutureGrid provides a user web portal and an important function of
this interface is to provide {\em resource configuration and load} information. This information describes the
clusters in FutureGrid, the nodes in the clusters, and how these nodes are assigned to different HPC and cloud
partitions. Users access this information for both long-term planning of which resources to use for their
experiments, short-term selection of which resources to use on a particular day, and similar tasks. The
FutureGrid portal provides information about the current resource configuration and load, but not historical
information.

Another type of information that is provided by the portal is {\em software and service descriptions}. These
descriptions include where software and services are located and how to access them. Similar to the resource
information, these descriptions also help users plan how to user FutureGrid. The portal and other users want
to examine the most recently published information about software and services and do not need to examine
older information.

A related type of information is {\em resource and service status} that describes whether resources and
services are operating correctly. Such status information is used by the providers of FutureGrid resources to
identify failures that need to be addressed and by users to determine which resources and services are
operating correctly at any given time. This information, failures in particular, must result in notifications
to those interested in them.

A final use case that is common to infrastructure projects is the ability to provide information about {\em
  resource usage}. This includes notifying users as their batch jobs and virtual machines change state as well
as accounting for resource use so that the portion of FutureGrid used by various projects over time can be
reported. This information should be available as both real-time updates and archived for post-analysis.

As an experimental infrastructure, the information system for FutureGrid also needs to support a few unique
use cases. One use case is the need to provide {\em detailed performance monitoring} information to
users. This includes dynamic information about the nodes such as processor load, memory usage, and disk I/O
operations as well as information about network traffic. This performance data lets users determine how their
experiments impact the FutureGrid infrastructure and is important input to many experiments. This information
will typically be observed in real time and archived in time-ordered streams.

A final use case is that FutureGrid must provide {\em federated information} for ease of use. The information
described above comes from a variety of sources in a variety of formats and it needs to be formatted and
provided in a unified manner.

\section{Monitoring Tools}
\label{sec-monitoring}

There is no single monitoring tool that provides all of the information needed by the use cases described in
the previous section so FutureGrid has deployed a number of specialized monitoring tools to satisfy them.
Each tool defines a schema for the information that it publishes and provides an interface to access the data
it collects.


Inca~\cite{Smallen:2007:UGM:1272680.1272687} is a monitoring framework designed to detect cyberinfrastructure
(CI) problems by executing periodic, automated, user-level probes of CI software and services.  Currently,
Inca runs 264 different tests at various frequencies to examine the components of the FutureGrid
infrastructure.  Inca monitoring results are published as eXtensible Markup Language (XML) and are accessible
through REST APIs and a Web interface.  Inca XML documents follow a self-defined ``reporter'' schema.


The Information Publishing Framework (IPF)~\cite{IPF-bitbucket,xsede13} is software developed as part of
XSEDE~\cite{XSEDE-web} to gather and publish information about clusters. It provides static and dynamic
information about a cluster including descriptions of the compute nodes, batch scheduling queues, and jobs
being managed by the batch scheduler or IaaS cloud software. This information is gathered by querying the
batch scheduler or cloud software managing the cluster. In addition, IPF monitors the batch scheduler and IaaS
infrastructure log files and publishes updates about jobs or virtual machines as they change state. IPF
represents this information using version 2 of the GLUE standard~\cite{GLUE2} published by the Open Grid Forum
(OGF). We have enhanced this software to produce information in the JSON in addition to XML.


perfSONAR~\cite{perfsonar} is an infrastructure for monitoring end-to-end network performance.  So far, the
FutureGrid perfSONAR deployment utilizes BWCTL to collect all-to-all Iperf bandwidth measurements.  We plan
to also include more frequent non-intrusive bandwidth measurements from either OWAMP or pingER.  perfSONAR's
network measurement results are published in XML and are accessible through a Web services API and a Web
interface.  perfSONAR's monitoring results follow the XML schema defined by the Network Measurement Working
Group of the OGF.


SNAPP~\cite{snapp} is a tool that collects high-performance, high-resolution SNMP data from network elements
and visualizes it.  SNAPP results are available in JSON format and are accessible through a REST interface and
Web pages.


Ganglia~\cite{ganglia} is a cluster monitoring tool that collects and reports detailed node data such as CPU,
memory, disk, and network usage.  Ganglia is installed on FutureGrid clusters and the data is collected
at a single server. Ganglia usage data is represented in XML and is accessible by connecting to a TCP port or
through its Web interface.  The XML documents follow an XML schema defined by Ganglia.


Finally, users can perform their own monitoring by integrating data gathering tools such as
NetLogger~\cite{netlogger} into the software and services they deploy as part of their experiments. This
allows users to record custom information that suits their specific needs. In addition, FutureGrid can use
NetLogger to instrument infrastructure services if users require performance data from those
services. NetLogger logs each event as a set of key value pairs.

\section{Design}
\label{sec-design}

The tools described in the previous section provide a great deal of information to FutureGrid users. However,
this information is delivered in different ways and in different formats and it is therefore difficult to use
together. One of the main goals of our information system is therefore to federate this information so that
it is easy to use.

A common way to do such federation is in a user interface such as a web portal. This approach is flexible
because the portal can be modified to incorporate new information sources and to present information in new
ways. However, this integration is only available to users of the portal; it isn't available to tools or users
of other interfaces, such as science gateways or from the command line. It also requires significant work by
the portal developers that isn't typically re-usable.

An alternate approach is to integrate information at a lower architectural level. A common way to do this is
with a centralized information storage system. Infrastructure projects have used relational databases, XML
databases, and the Lightweight Directory Access Protocol (LDAP) to do this. It can be a complex task to
configure and manage a shared storage system containing information about a distributed infrastructure,
but this effort reduces the effort needed on other parts of the project.

\begin{figure*}
\begin{center}
\centerline{\includegraphics[width=4.5in]{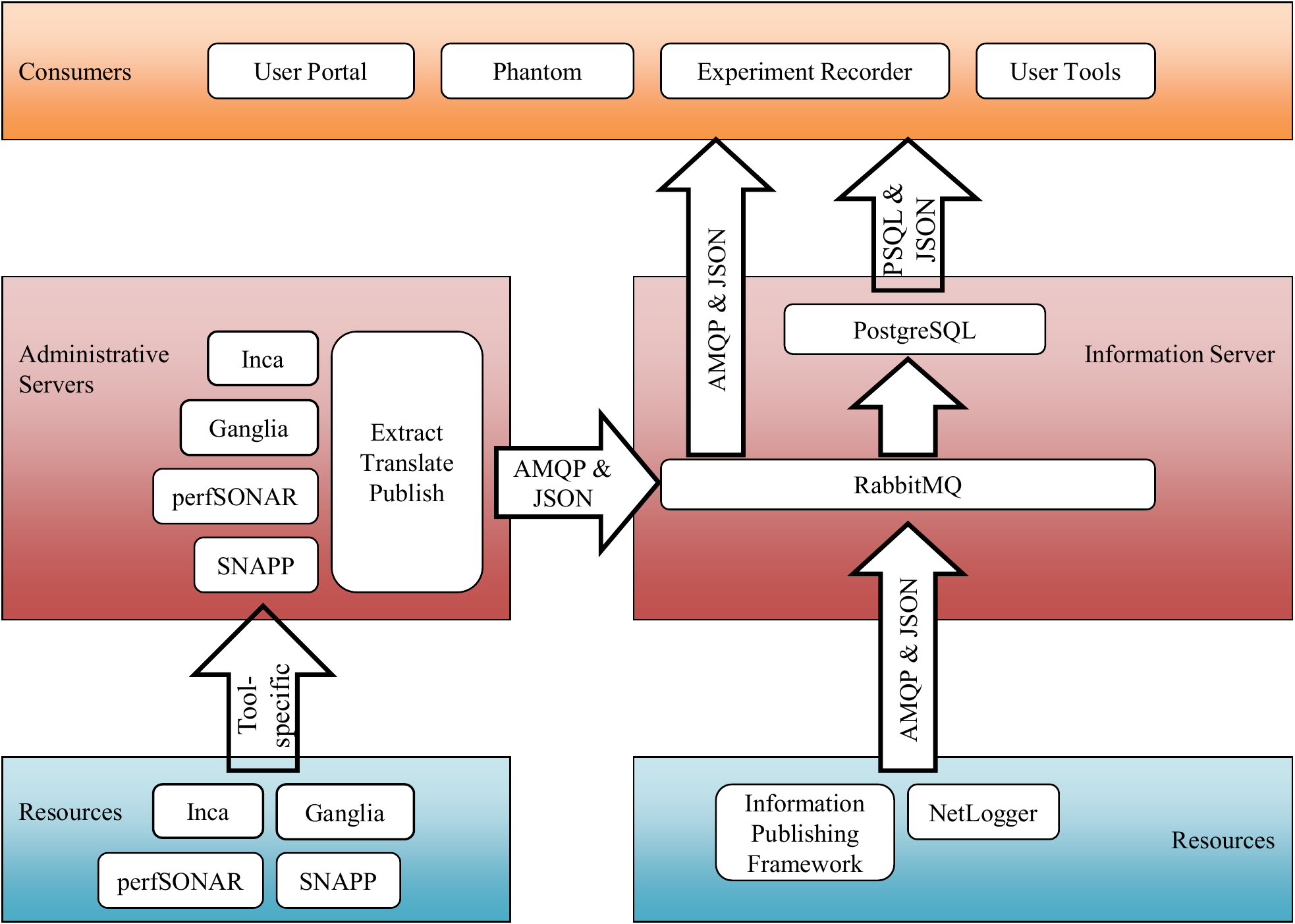}}
\end{center}
\vspace{-0.25in}
\caption{The FutureGrid monitoring information system.}
\label{fig-futuregrid}
\end{figure*}

However, a database model may not be the best one to use as the foundation since a publish/subscribe model is
a more natural fit for many use cases, such as the ones described in Section~\ref{sec-usecases}, where
consumers want to act on updated information as it arrives.  A publish/subscribe model is also a good fit
where information needs to be transformed.  We therefore developed and deployed the design shown in
Figure~\ref{fig-futuregrid}.

The lowest layer of this design consists of the monitoring tools described previously. Many of these tools
have an Extract, Transform, and Publish (ETP) service layered on top of them to integrate them into the
information system. As you would expect, these services extract information from the monitoring tools in
tool-specific ways, transform that information into a common representation language, and publish the
transformed information into the information system.

Our first design decision was to use JSON~\cite{json-web} as the common representation language. The
monitoring tools deployed on FutureGrid produce data in either XML or JSON so it was natural to select one of
those two languages. We selected the JSON format because it is sufficiently expressive, it is very easy to
parse and generate programmatically, and it is easy to read by a person. There are also JSON libraries
available for many different programming languages. The selection of a common representation language requires
that we transform the XML documents produced by some of the monitoring tools into JSON documents.

This translation of XML documents to JSON documents is one of the tasks of the ETP services where these
services create JSON documents that resemble the original documents as closely as possible. We chose not to
enforce common naming or data formats across documents because this would make it more difficult for users
that are already familiar with the documents produced by specific monitoring tools.

One of our main design decisions was to use publish/subscribe messaging to distribute information. This model
is a very good fit for publishing monitoring information where new versions of information are constantly
being generated (e.g. the current state of a resource or service). This model is also a good fit to many
consumers of monitoring information that wish to be updated with the most recent information, to watch for
exceptional information, or to log information in time order. We decided to use a standard messaging protocol
called the Advanced Message Queuing Protocol (AMQP)~\cite{AMQP-0.9.1,AMQP-web}. There are several
production-quality messaging services that implement this standard and also a wide variety of client
libraries. Selecting a standard protocol allows us to more easily switch to a different client library or
messaging service if we encounter problems with specific software. The ETP services publish transformed
information using AMQP for those monitoring tools.

There two monitoring tools that do not need an ETP service. IPF supports JSON and one of the publishing
options it provides is via AMQP. The resource, queue, and job/virtual machine descriptions provided by IPF in
the GLUE2 format are currently available to users.  The other monitoring tool that does not have an ETP
service is NetLogger.  NetLogger supports publishing of performance log messages via AMQP, but the messages
are formatted as key value pairs, rather than JSON. Our goal is to allow FutureGrid users to embed NetLogger
into their programs so that they can monitor custom performance information. An ETP service isn't the best
approach in this case where NetLogger will be used in a dynamic number of locations by a variety of users. The
approach we have under development is to provide a version of NetLogger to FutureGrid users that formats log
messages as JSON objects - a very simple format change.

For the other monitoring tools, the ETP services are configured in the following ways:
\begin{itemize}
\item {\em Inca} (available): ETP service runs on the centralized Inca storage server as a ``depot filter'';
  each time a new Inca report is received, the incoming report XML document is transformed to JSON and is
  published to the messaging service.
\item {\em PerfSONAR} (under development): ETP service runs on the perfSONAR data server and queries the Web Services
  API once a minute, determines which network links have updated information, translates those XML documents
  to JSON, and publishes the JSON documents to the messaging service.
\item {\em SNAPP} (under development): ETP service runs on the SNAPP data server and queries the SNAPP rest service
  several times a minutes, determines which network links have updated information and publishes the JSON
  documents to the messaging service.
\item {\em Ganglia} (available): ETP service runs on the virtual machine that collects all Ganglia
  information for FutureGrid. This service queries the Ganglia gmetad several times a minute, determines which
  machines have updated information, translates those XML documents to JSON, and publishes the JSON documents
  to the messaging service.
\end{itemize}

FutureGrid selected RabbitMQ~\cite{RabbitMQ-web} as its AMQP messaging service. RabbitMQ is a
production-quality messaging service that provides mechanisms for scalability and fault tolerance and has been
shown to have high performance~\cite{PubSub-TG11,RabbitMQ-perf}. This is an important feature for FutureGrid
given our higher level of information gathering as well as our desire to support user-generated custom
performance information.  RabbitMQ is configured so that different types of information are delivered to
pre-defined logical locations (exchanges) and users subscribe for this information at those locations. In
addition, each message has a tag (routing key) that follows a pre-defined format that users can filter on.

While many consumers of FutureGrid monitoring information prefer to receive that information via messaging, we
also wanted to provide a service where this information can be stored and searched. After investigating
several NoSQL storage technologies that were appropriate for storing JSON documents, we decided to instead use
the PostgreSQL~\cite{PostgreSQL-web} database. PostgreSQL supported the highest update rate on a single server
of the technologies we tested and the PostgreSQL developers recently added a JSON type and operations. This
JSON support allows users to easily define searches made over JSON data stored in columns and while inserting
JSON documents, we can also extract a few key pieces of information (such as resource and service names) and
include them in other columns of the same table for traditional SQL searches. This approach provides a hybrid
of relational, key/value, and document-oriented models and provides some of the benefits of all of them.

\section{Performance}
\label{sec-performance}

Before finalizing the design of this information system, we performed a set of performance experiments to
ensure that the design can satisfy the relatively high demands placed on it by FutureGrid. These experiments
consist of a set of experiments to determine the throughput limits of our messaging and database services and
a set of experiments that emulate FutureGrid to confirm that FutureGrid will operate within these limits. In
addition, we performed emulations on an experimental infrastructure twice the size of FutureGrid to ensure
that our design will scale to potential future needs. Finally, since we believe that our approach is a good
one for distributed scientific infrastructures, we performed emulations of XSEDE, the Open Science Grid, and
infrastructures twice their current size to demonstrate that this design would also satisfy their needs.

\subsection{Experimental Environment}

\begin{figure}
\begin{center}
\centerline{\includegraphics[width=3.25in]{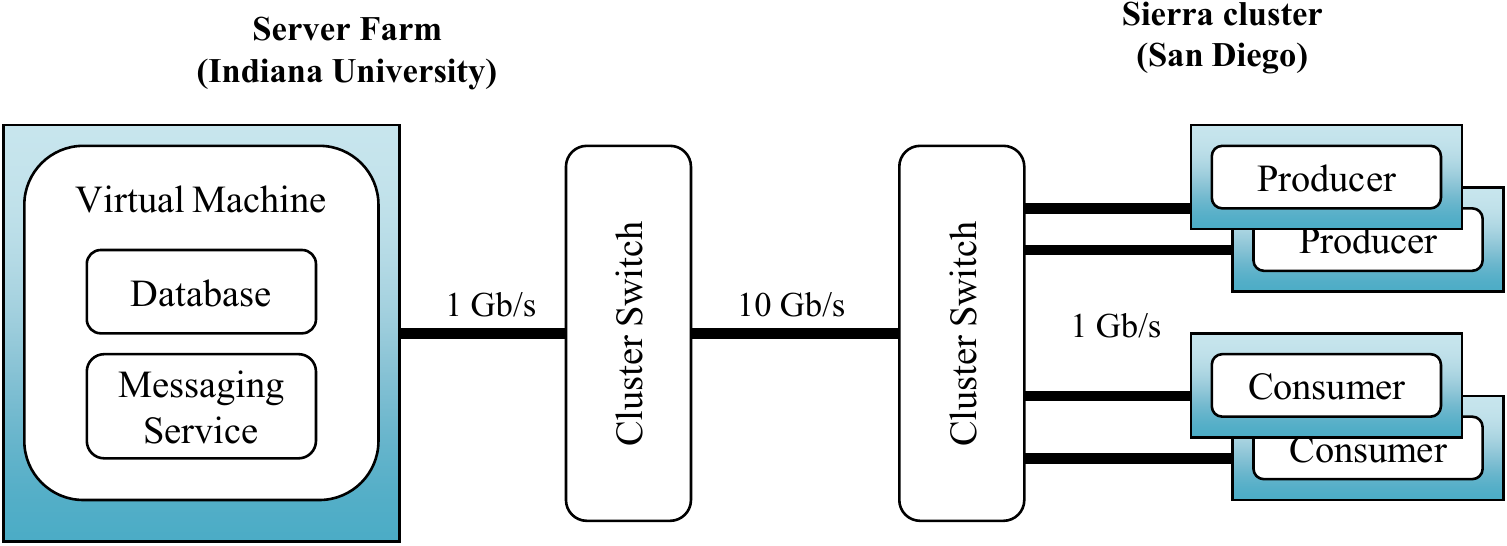}}
\end{center}
\vspace{-0.25in}
\caption{Overview of the experimental environment.}
\label{fig-environment}
\end{figure}

We executed our performance experiments on FutureGrid using the experimental environment shown in
Figure~\ref{fig-environment}. The messaging service or database is running in a virtual machine at Indiana
University - the same virtual machine as where these services are currently in production. This KVM virtual
machine has four 2.4 GHz virtual CPUs and 4GB of memory and a virtio network interface.

A set of Publishers and Subscribers are running on compute nodes of the FutureGrid Sierra cluster at the San
Diego Supercomputing Center. The programs emulate the actions of various monitoring tools that produce
information and various consumers of that information in a cyberinfrastructure. The producers and consumers
that interact with RabbitMQ are written in Java and use version 3.1.1 of the RabbitMQ Java client library. The
producers and consumers that interact with PostgreSQL are also written in Java, but use the PostgreSQL 9.2
JDBC driver. Each of the Sierra compute nodes has two Intel L5420 processors (total of 8 cores) running at 2.5
GHz with 32GB of memory.

The compute nodes on both clusters are connected to their local cluster Ethernet switch at 1 Gbit/sec. There
is a 10 Gbit/sec network path between the two cluster switches via the FutureGrid network.

\subsection{Throughput}
\label{sec-througput}

Our first experiments examine the maximum throughput that can be attained when using RabbitMQ and PostgreSQL
on the FutureGrid infrastructure. RabbitMQ is measured by sending messages of various sizes as fast
as possible from $N$ producers to $N$ consumers, with each consumer subscribed to the messages published
by a single producer. RabbitMQ is configured using the default options and communication with clients is
unencrypted.  The messaging client libraries are also used with their default configurations.  There are
optimizations that could improve performance (such as not acknowledging every message individually), but we
used the default configurations because that is likely what many producers and consumers would do.

PostgreSQL throughput is measured by having $N$ producers update records into a table as fast as possible
where the records consist of an identifier, a producer-specific key, and a block of text of various sizes
while $N$ subscribers select a single record by key as fast as they can.

\begin{figure}
\begin{center}
\centerline{\includegraphics[width=3.25in]{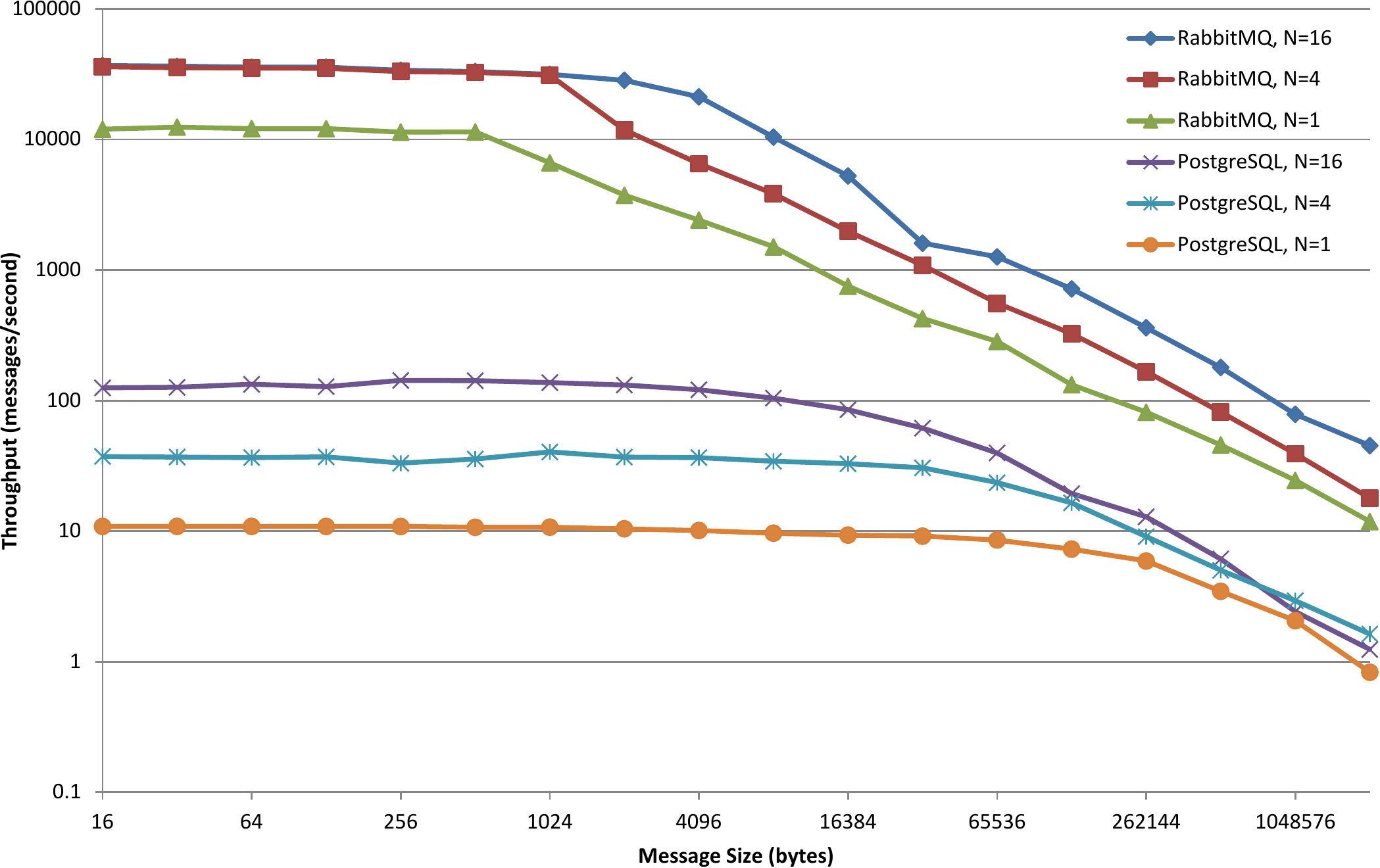}}
\end{center}
\vspace{-0.25in}
\caption{Throughput of $N$ producers to $N$ consumers.}
\label{fig-throughput}
\end{figure}

The throughput achieved for these experiments is contained in Figure~\ref{fig-throughput} and shows that
RabbitMQ can deliver over 36,000 small messages per second in our environment. The data also shows that
RabbitMQ provides over two orders of magnitude higher throughput than PostgreSQL. This result is not
surprising given that as a database, PostgreSQL must perform more complex computations and disk accesses than
a publish/subscribe messaging service.

One interesting thing to note is that increasing the number of producer and consumer threads significantly
increases the throughput of PostgreSQL. We will take advantage of this fact to meet our update rate goals for
FutureGrid. Increasing the number of threads also increases the throughput of RabbitMQ, but not as
significantly.

While PostgreSQL provides significantly lower throughput than RabbitMQ, it is the best data storage/search
approach that we found for our needs since the NoSQL approaches we examined are targeted to different problems
than ours. For example, we found that a single writer could only perform approximately 10 writes per second to
a MongoDB document-oriented database and we did not want to scale horizontally to additional servers to
improve performance. CouchDB stores JSON documents and supports higher write rates than MongoDB, but it is
meant for incremental additions and doesn't fully delete documents (we do not wish to permanently store
high-volume data such as that produced by Ganglia or SNAPP). Couchbase is another JSON document database with a
goal of providing efficient access to large volumes of data, but it keeps all keys in memory and this isn't a
good fit to our resource-constrained deployment.

\subsection{FutureGrid Emulation}
\label{sec-emulation}

To emulate FutureGrid, we characterize the features that impact the publication of the monitoring information
described in Section~\ref{sec-monitoring}.  The important characteristics are:
\begin{itemize}
\item The number of partitions. On FutureGrid, many clusters are operated as two or more partitions with
  different resource management systems per partition. IPF therefore publishes GLUE2 information about each
  partition. We emulate IPF publishing partition configuration/load and queue snapshots every 2 minutes.
\item The number of jobs being managed at any time. This characteristic affects the size of the GLUE2 queue
  documents.
\item The job throughput. A GLUE2 job document is generated for each job state change. Information about each
  job published 3 times during the lifetime of a job (submit, start, end).
\item The number of services. Some services are associated with each cluster and others are associated with
  FutureGrid as a whole. In either case, each service has several Inca tests that are run periodically against
  it every 15 to 120 minutes.
\item The number of nodes. Ganglia gathers 37 metrics on each node and new metrics for a node are available
  every 15 seconds.
\item The number of network links. The SNAPP measurements are determined by this and provide network traffic
  data for each link every 10 seconds. This also affects the number of bandwidth test results reported by
  perfSONAR, which are run every two hours.
\end{itemize}

To approximate consumers of information, we emulate the following subscribers:
\begin{itemize}
\item Information databases that want all information.
\item User web portal that also wants all information.
\item Accounting systems that wants GLUE 2 job updates.
\item Metaschedulers that want GLUE 2 snapshots and job updates.
\item Monitoring system that wants all Inca, perfSONAR, SNAPP, and Ganglia information.
\item Science Gateways that want GLUE 2 snapshots, GLUE 2 job updates, and Inca test results.
\end{itemize}

The values that we choose for these characteristics (based on observations of FutureGrid) are shown in the
leftmost column of Table~\ref{tab-config}. We performed an emulation with these characteristics using
multi-threaded messaging publishers and subscribers in the environment shown in
Figure~\ref{fig-environment}. The results are shown in the top row of Table~\ref{tab-messaging-results}.  For
FutureGrid, about 41 messages per second are published but since the size of the messages are relatively small
(approximately 2 KB), the data bandwidth is a low 0.09 MB/sec.  The number of consumers that we emulate,
described above, multiply these characteristics significantly on the delivery side where 1,101 messages a
second are delivered to consumers at a data rate of 2.38 megabytes per second. Since our throughput
experiments of Figure~\ref{fig-throughput} show that even one publisher and one subscriber can transmit 2 KB
messages at over 3,700 per second and sixteen publishers and subscribers can transmit 2 KB at over 28,300 per
second, the messaging implementation in our information system can therefore easily handle the messaging
volume we need for FutureGrid.

To ensure that our information system has the capacity to grow with FutureGrid, we also preformed emulations
of FutureGrid at twice its current size. The third column of Table~\ref{tab-config} shows this configuration
and the second row of data in Table~\ref{tab-messaging-results} shows the performance results. This doubling
in size did not impact the time to publish messages and the throughput and bandwidth from the publisher
doubled. The throughput and bandwidth on the consumer side tripled, but again the 3,254 messages a second
delivered is comfortably under the 28,300 per second observed in Figure~\ref{fig-throughput}.

\begin{table*}
\begin{center}
\caption{Emulation configurations.}
\begin{tabular}{|c||c|c||c|c|c|c|}
\hline
                   & FutureGrid & FutureGridx2 & XSEDE & XSEDEx2 &   OSG & OSGx2 \\ \hline \hline
Partitions         &         14 &           28 &    13 &      26 &   200 &   400 \\ \hline
Simultaneous Jobs  &        477 &          954 &  6600 &   13200 & 42300 & 84600 \\ \hline
Jobs per Hour      &         78 &          154 &  1090 &    2169 & 21254 & 42455 \\ \hline
Services           &         77 &          144 &   260 &     520 &  4000 &  8000 \\ \hline
Nodes              &        608 &         1216 &   N/A &     N/A &   N/A &   N/A \\ \hline
Network Links      &          6 &           12 &   N/A &     N/A &   N/A &   N/A \\ \hline \hline
Info. databases    &          1 &            1 &     1 &       1 &     1 &     1 \\ \hline
Web portals        &          1 &            1 &     1 &       1 &     1 &     1 \\ \hline
Accounting systems &          1 &            1 &     1 &       1 &     1 &     1 \\ \hline
Metaschedulers     &          1 &            2 &     2 &       4 &     2 &     4 \\ \hline
Monitoring systems &          2 &            4 &     1 &       1 &     1 &     1 \\ \hline
Science Gateways   &          0 &            0 &    10 &      20 &    20 &    40 \\ \hline
\end{tabular}
\label{tab-config}
\end{center}
\end{table*}

\begin{table*}
\begin{center}
\caption{Summary of messaging experiments.}
\begin{tabular}{|c||c|c|c|c||c|c|}
\hline
           & \multicolumn{4}{c||}{Published} & \multicolumn{2}{c|}{Consumed} \\
           &    Average   &   Average    &             &           &            &           \\
           & Publish Time & Message Size &  Throughput & Bandwidth & Throughput & Bandwidth \\
Experiment &     (msec)   &    (bytes)   &  (msg/sec)  & (MB/sec)  & (msg/sec) &  (MB/sec)  \\
\hline \hline
  FutureGrid &  0.09 &  2,175 & 41.46 & 0.09 &  1,101 &  2.38 \\ \hline
FutureGridx2 &  0.08 &  2,174 & 80.78 & 0.18 &  3,254 &  7.01 \\ \hline
       XSEDE & 29.27 & 10,517 &  1.65 & 0.02 &    163 &  1.65 \\ \hline
     XSEDEx2 & 26.08 & 10,443 &  3.32 & 0.03 &    593 &  6.15 \\ \hline
         OSG &  4.34 &  4,367 & 27.26 & 0.12 &  4,572 & 19.54 \\ \hline
       OSGx2 &  3.79 &  4,394 & 54.39 & 0.24 & 17,222 & 75.08  \\ \hline
\end{tabular}
\label{tab-messaging-results}
\end{center}
\end{table*}

We also performed emulations of the PostgreSQL database to determine if it can handle the amount of data that
flows through the messaging service.  In this emulation, the database only stores the most recently received
information about partition configuration, load, and usage. The database does store all job/virtual machine,
Inca, Ganglia, SNAPP, and perfSONAR received during the hour-long duration of the emulation.  These emulations
were configured to perform inserts/updates for all monitoring information in the same way that our messaging
emulations published a message for each piece of updated monitoring information. One significant difference in
these experiments are that the consumers were configured to select on the monitoring information they are
interested in once a minute. This emulates a user or tool querying for current information at a relatively
frequent rate. One effect of this is that for high rate data streams, such as Ganglia, the consumers in these
emulations do not see all of the values.

The results of these emulations are shown in Table~\ref{tab-database-results} and show that the average update
size, throughput, and bandwidth match the values seen for messaging. This was accomplished because 16 threads
were used to emulate producers writing into the database. One of the results of the data in
Figure~\ref{fig-throughput} is that multiple threads can be used to scale the throughput to the database and
16 threads provides an insert/update rate of over 133 per second. This rate is sufficient for the 41 messages
per second published in the FutureGrid emulation and the 81 messages per second published in the FutureGridx2
emulation.

\begin{table*}
\begin{center}
\caption{Summary of database experiments.}
\begin{tabular}{|c||c|c|c|c||c|}
\hline
           & \multicolumn{4}{c||}{Published} & \multicolumn{1}{c|}{Consumed} \\
           &    Average  &   Average   &               &           &               \\
           & Update Time & Update Size &  Throughput   & Bandwidth &   Throughput  \\
Experiment &     (msec)  &    (bytes)  & (updates/sec) & (MB/sec)  & (selects/sec) \\
\hline \hline
  FutureGrid & 127.74 &  2,175 & 41.46 & 0.09 &  3.26 \\ \hline
FutureGridx2 & 120.44 &  2,174 & 80.78 & 0.18 &  4.91 \\ \hline
       XSEDE & 225.07 & 10,517 &  1.65 & 0.02 &  7.22 \\ \hline
     XSEDEx2 & 228.20 & 10,443 &  3.32 & 0.03 & 12.84 \\ \hline
         OSG & 202.70 &  4,367 & 27.26 & 0.12 & 10.38 \\ \hline
       OSGx2 & 218.02 &  4,394 & 54.39 & 0.24 & 22.24 \\ \hline
\end{tabular}
\label{tab-database-results}
\end{center}
\end{table*}

\subsection{Scientific Cyberinfrastructure Emulation}

A number of communities have deployed large-scale distributed systems in support of science around the
world. In the United States, the National Science Foundation adopted the term ``cyberinfrastructure'' for such
distributed systems and supported the deployment of several of them. FutureGrid is a small example of such
an infrastructure while eXtreme Science and Engineering Discovery Environment (XSEDE) and the Open Science
Grid (OSG) are much larger examples.

Cyberinfrastructures require a certain amount of monitoring of configuration, operational status, and resource
load. This information is used by the engineers operating the infrastructure and by scientists using the
infrastructure. In this section, we examine how well our information system can provide this functionality for
such cyberinfrastructures by emulating the information production and consumption of such infrastructures.

We believe that our information system approach is also a good one for such cyberinfrastructures for many of
the same reasons it works well on FutureGrid. Real-time delivery of information is useful for monitoring the
resources, services, and jobs in cyberinfrastructures and a publish/subscribe messaging architecture provides
this functionality. The use of a single language to represent information makes information easier to use and
JSON is a good choice for this language because it is easy to use programmatically and is also readable by
humans. Finally, also providing a mechanism so that users can search recent JSON information documents is also
valuable in some use cases.

The main issue to address is whether this information system approach can satisfy the needs of large-scale
cyberinfrastructures such as XSEDE and OSG. We address this question by emulating these infrastructures.
Table~\ref{tab-config} provides the parameters we use to drive emulations of XSEDE, OSG, and for emulations of
these infrastructures if they were two times their current size.  For these emulations, we do not emulate
Ganglia, perfSONAR, or SNAPP measurements since this information is specific to experimental infrastructures
such as FutureGrid. The number of nodes and links in XSEDE and OSG is therefore not applicable. The other
parameters we derived by first-hand observation of XSEDE and by examining the monitoring information provided
by OSG~\cite{OSG-web}. One thing to note is that XSEDE is made up of fewer, large systems while OSG is made up
of many, smaller systems. This different impact the amount of monitoring information that they generate.

The results of these emulations are shown in Table~\ref{tab-messaging-results} and
Table~\ref{tab-database-results}. As you can see, the time to publish messages or update rows is higher than
for the FutureGrid experiments. One reason for this is that the average message size is larger for XSEDE and
OSG messages. However, this does not account for all of the difference in publish time between FutureGrid and
OSG for messaging. The Ganglia messages in the FutureGrid experiments are published to RabbitMQ very quickly
(approximately 0.05ms for each) and these messages make up almost 98\% of the published FutureGrid
messages. We believe that the small size and consistency of routing keys of these messages let RabbitMQ
optimizing the routing of them.

The volume of Ganglia messages also results in the relatively small FutureGrid infrastructure publishing more
messages per second than even OSG. However, the number of consumers for OSG results in more messages being
delivered. In fact, for the OSGx2 experiment, 17,222 messages are delivered per second. This is relatively
close to the approximately 21,200 messages per second of size 4 kilobytes that our throughput experiments in
Figure~\ref{fig-throughput} suggest that RabbitMQ can deliver. One way to handle a situation with high numbers
of message deliveries for each message received like this is to deploy multiple distributed messaging servers
and have consumers subscribe to different ones. This spreads the work of delivering messages across multiple
servers on different networks.

\subsection{Custom Experiment Information}

In addition to providing detailed performance information to FutureGrid users, one of our goals is to let
users use the information system to publish custom information while their experiments run. One way of doing
this is for users to instrument their software and services using NetLogger~\cite{netlogger}.

We expect that these custom measurements will typically be small pieces of performance information or
notifications that events have taken place. Figure~\ref{fig-throughput} shows that for messages containing 2
KB of information, over 28,000 messages per second can be handled. Table~\ref{tab-messaging-results} shows
that shared FutureGrid monitoring tools may publish about 1,101 messages per second of slightly over 2 KB.
Therefore, approximately 96\% of the messaging capacity is available to FutureGrid users for custom experiment
information. This table also shows that even expanding FutureGrid by a factor of two would still leave over
88\% of the messaging capacity available for custom experiment information.


\section{Related Work}
\label{sec-relatedwork}

A number of information systems have been proposed and deployed on scientific infrastructures. For example,
early versions of the Globus software included a Metacomputing Directory Service (MDS)~\cite{Globus-2} for
storing information in a hierarchical directory service accessed via the Lightweight Directory Access Protocol
(LDAP). The Globus project then transitioned to a MDS based on web service (SOAP/WSDL)
technologies~\cite{MDS4}.

The Condor system contains a Collector service that contains classads that describe hosts. These descriptions
contain standard information, but can also include custom information. This service is uses as part of Condor
matchmaking~\cite{raman-hpdc7}.

XSEDE~\cite{XSEDE-web} is an NSF infrastructure that provides a set of large resources for scientific
simulation and analysis. XSEDE is a continuation of TeraGrid~\cite{teragrid} and is currently using the TeraGrid
information services. There are a variety of TeraGrid/XSEDE
information subsystems~\cite{TeraGrid-IIS} and these subsystems are only partially integrated. The Integrated
Information Service (IIS) is implemented using the Globus WS-MDS~\cite{MDS4}, a set of
distributed services that support reading and writing of XML documents via web service interfaces. The IIS
contains information about resource configuration, resource load, and the software and services deployed on
systems. There is a separate XSEDE database contains information about users, allocations and jobs run against
allocations. There is also a separate Resource Description Repository that contains manually-entered
information about XSEDE resources. The XSEDE user portal integrates these various sources of information.

The Open Science Grid (OSG) is a consortium of eighty sites that advances science through open distributed
computing~\cite{OSG}. It originates from the high-energy physics community and now supports a number of other
scientific communities.  OSG uses an older version of GLUE to publish resource and software information using
a Condor-based Resource Selection Service (ReSS) service~\cite{ReSS}.  The GLUE data is collected centrally
using a LDAP-based server called the Berkeley Database Information Index (BDII)~\cite{OSG-web}.  For
monitoring, OSG utilizes the Resource and Service Validation (RSV) software~\cite{RSV} consisting of a client
that executes a number of tests and publishes it to a centralized accounting service called
Gratia~\cite{Gratia}.  All monitoring and information services are unified under a web portal called
MyOSG~\cite{myosg}, but they do not provide unified programmatic interfaces as described in this paper.

The Partnership for Advanced Computing in Europe (PRACE) spans twenty-four countries to provide a
supercomputing infrastructure for Europe~\cite{PRACE}.  Like FutureGrid, PRACE also utilizes Inca to verify
its software infrastructure, the PRACE Common Production Environment~\cite{PRACEOPS} as well as perfSONAR for
network monitoring~\cite{PRACEOPS}.  Like OSG, their web pages do not indicate a project to unify the
monitoring tools under a single service.

The European Grid Infrastructure (EGI) is a federation of approximately forty resource providers to deliver a
sustainable, integrated and secure computing services to European researchers and their international
partners~\cite{EGI}.  For monitoring, EGI uses Gstat~\cite{Gstat}, a monitoring solution built on top of
Nagios~\cite{Nagios}. EGI has deployed several instances of the ActiveMQ message broker and is experimenting
with using messaging in their infrastructure. One example is publishing status information gathered by tools
such as Nagios to these brokers.


\section{Conclusions and Future Work}
\label{sec-conclusions}

Distributed infrastructures are complex systems that must provide monitoring information to users and to the
personnel managing the infrastructure. FutureGrid is unique because it is an experimental testbed and provides
a great deal of performance information in addition to the resource, service, and software information
provided by typical distributed infrastructures.

This heterogeneous and real-time information is gathered by a variety of monitoring tools and needs to be
federated and provided to users and managers in an efficient and easy to use manner.  The information system
described in this paper federates information at a low architectural level via publish/subscribe messaging and
a common representation language.  In addition, the information system also includes a database to store
recently generated information in a searchable manner.

We selected JSON as our common representation language since several of our monitoring tools already supported
it.  In addition, we selected JSON because it is sufficiently expressive, very easy to parse and generate
programmatically, and easy to read by a person.  We found it to be straightforward to translate all of our
monitoring information to JSON.

We found that publish/subscribe messaging is an effective model to use in an information system. This model
matches the publishing mechanisms of our monitoring tools and is the preferred delivery model for many of our
use cases. Furthermore, our performance results indicate that the message service we selected, RabbitMQ, can
handle a very large volume of messages which allows for the possibility of FutureGrid users generating their
own custom performance information and publishing it to the messaging system.

Since a messaging service does not support searching over stored information, our information system also
includes a PostgreSQL database to support such functions. In addition to having better performance than other
data stores that we tested, PostgreSQL has recently included a JSON data type and operations that act on the
information in JSON documents stored in columns. This lets us use PostgreSQL as a hybrid relational and
document-oriented database and provide a very flexible information storage and search functionality. Our
performance experiments found that PostgreSQL provides a significantly lower throughput than RabbitMQ, but
that with the use of multiple threads updating PostgreSQL information, it can keep up with the amount
of data generated by FutureGrid and an expanded version of FutureGrid.

Finally, we believe that our information system approach can be applied to large distributed scientific
infrastructures such as XSEDE and OSG. Our design provides the functionality needed to satisfy typical use
cases of such infrastructures and our performance experiments show that our implementation has the capacity to
support the current size of XSEDE and OSG as well as expanded versions of these infrastructures.

The next step of this work is to complete the last few components so that all FutureGrid monitoring
information is available in this information system. After that is complete, we will investigate providing
tools to FutureGrid users so that they can publish their own custom information into the information system.

\section*{Acknowledgment}

This work was supported by the National Science Foundation under grant 0910812.

\bibliographystyle{IEEEtran}
\bibliography{references}

\end{document}